\pdfoutput=1

\documentclass[11pt]{article}

\usepackage{amsmath} 
\usepackage{amssymb}
\usepackage{booktabs} 
\usepackage{multirow}
\usepackage{makecell}
\usepackage{xcolor}

\usepackage[final]{acl}

\usepackage{times}
\usepackage{latexsym}

\usepackage[T1]{fontenc}

\usepackage[utf8]{inputenc}

\usepackage{microtype}

\usepackage{inconsolata}

\usepackage{graphicx}

\title{Improving Table Retrieval with Question Generation from Partial Tables}

\author{Hsing-Ping Liang, Che-Wei Chang, Yao-Chung Fan$^*$ \\
  Department of Computer Science and Engineering, \\National Chung Hsing University, Taiwan \\
  \texttt{yfan@nchu.edu.tw}
}

\begin{document}
\maketitle
\begin{abstract}
Recent advances in open-domain question answering over tables have widely adopted large language models (LLMs) under the Retriever-Reader architecture. Prior works have effectively leveraged LLMs to tackle the complex reasoning demands of the Reader component, such as text-to-text, text-to-SQL, and multi-hop reasoning. In contrast, the Retriever component has primarily focused on optimizing the query representation—training retrievers to retrieve relevant tables based on questions, or to select keywords from questions for matching table segments. However, little attention has been given to enhancing how tables themselves are represented in embedding space to better align with questions. To address this, we propose QGpT (Question Generation from Partial Tables), a simple yet effective method that uses an LLM to generate synthetic questions based on small portions of a table. These questions are generated to simulate how a user might query the content of the table currently under consideration. The generated questions are then jointly embedded with the partial table segments used for generation, enhancing semantic alignment with user queries. Without the need to embed entire tables, our method significantly improves retrieval performance across multiple benchmarks for both dense and late-interaction retrievers.\footnote{The code and reconstructed corpora are available at \url{https://github.com/cc3374twa/QGpT}} 

\end{abstract}

\section{Introduction}
\begin{figure*}[t]
  \includegraphics[width=\linewidth]{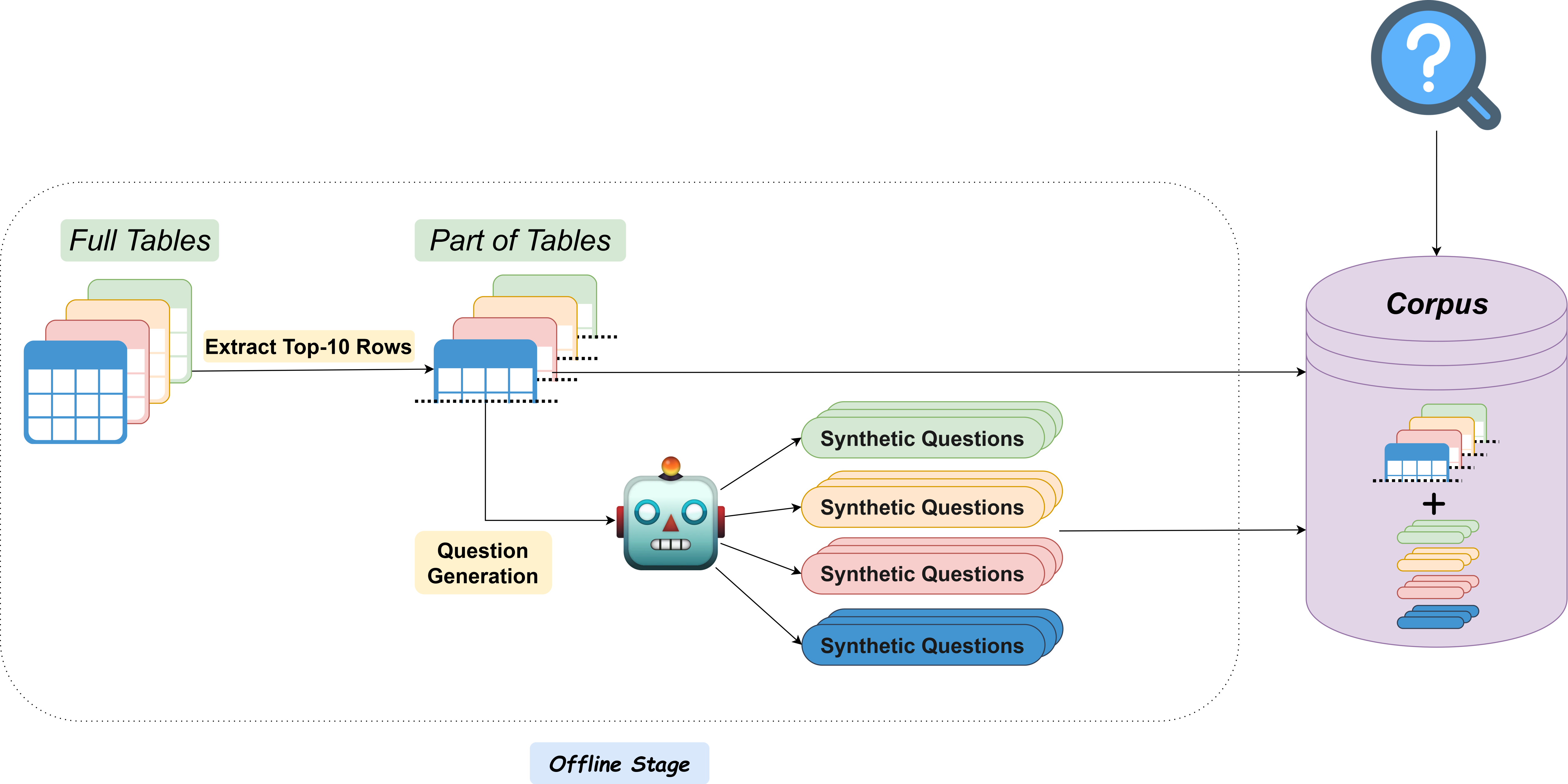}
  \caption{The \textbf{QGpT} pipeline: In the offline stage, top-10 rows from full tables are used to generate synthetic questions via LLM. The questions and table snippets are embedded and stored in the corpus, enhancing retrieval through semantic alignment without encoding full tables.}
  \label{fig:Overview}
\end{figure*}
\begin{figure*}[t]
  \includegraphics[width=\linewidth]{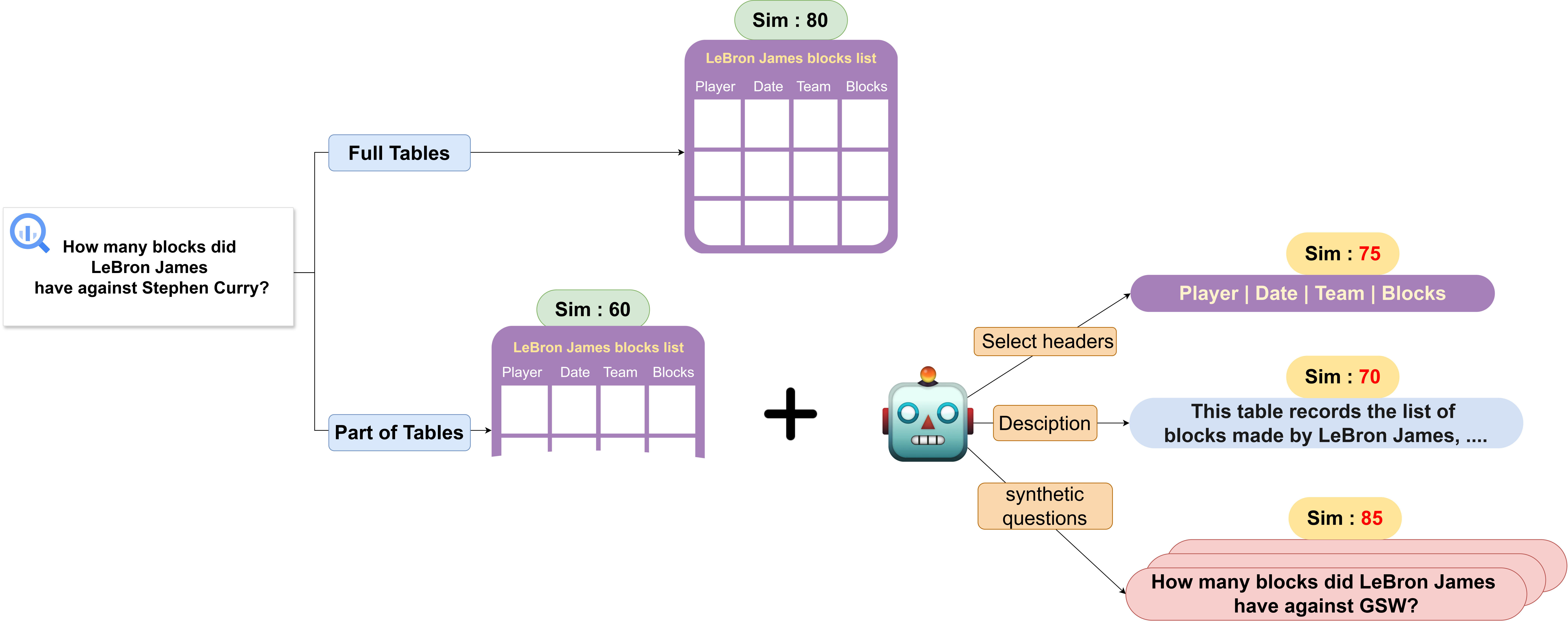}
  \caption{An illustration of three strategies for enriching truncated tables to enhance semantic alignment with the input query: selecting headers, generating table descriptions, and generating synthetic questions.}
  \label{fig:Table-representation}
\end{figure*}
Table-based question answering (Table-QA) has drawn increasing attention in recent years, beginning with early works on Wiki-style tables~\citep{WikiTableQuestions,WikiSQL}. These early Table-QA tasks typically operate under the assumption that the relevant table is provided alongside the question. While this assumption simplifies the problem and allows a focused evaluation of reasoning over structured data, it fails to reflect the challenges of real-world open-domain usage scenarios.

In practical settings, users do not typically specify which table to consult. Instead, they ask questions in natural language, and the system must first determine which tables might contain the relevant information before reasoning. The assumption of a known target table is thus insufficient in many realistic applications, where the answer may reside in any one of potentially thousands of tables within a large corpus.

To address this gap, recent research has moved toward integrating table retrieval into the QA pipeline. This shift is marked by the adoption of Retriever-Reader architectures~\citep{DRQA}, where a retriever component is responsible for identifying candidate tables, followed by a reader module that performs QA over the retrieved tables. 

As a pioneering study, NQ-TABLES~\citep{NQ-TABLES} provides naturally phrased questions and their corresponding answer tables extracted from the Natural Questions dataset~\citep{NQ}, and transforms TAPAS~\citep{TAPAS}, a BERT-based table reader, into a dense table retriever (DTR) by adopting the DPR~\citep{DPR} training paradigm. 

Yet another study named CLTR \citep{E2E-WTQ} integrates the RCI (Row-Column Intersection;~\citealp{RCI}) model with BM25~\citep{bm25} for table retrieval, and introduces the E2E-WTQ dataset for RCI retriever fine-tuning. 

However, studies such as NQ-TABLES and CLTR operate under the assumption that the answer to a question resides in a specific cell within a single, relatively short table. While this setting facilitates controlled evaluation, it significantly limits the generalizability of these approaches to real-world applications, where answering questions often requires reasoning over multiple or lengthy tables.

Therefore, MMQA~\citep{MMQA} extend the Table-QA paradigm by introducing complex reasoning tasks that integrate question answering with Text-to-SQL, using the Spider dataset~\citep{spider} as a foundation. They propose a \textit{multi-table retrieval} (MTR) setting, in which a single question may require retrieving and reasoning over multiple relevant tables. Their approach decomposes the original query into sub-queries, enabling independent retrieval and re-ranking of candidate tables.

Building on the idea of query decomposition, TableRAG~\citep{TableRAG} adopts a similar strategy by splitting questions into schema-level and cell-level components and performing separate retrieval for each. Additionally, it segments tables into schema and content parts, allowing scalable retrieval across corpora with millions of cells, and focuses on retrieving localized table segments to improve efficiency and accuracy.



In summary, Table-QA has evolved from simple single-table, short-answer settings~\citep{WikiTableQuestions,WikiSQL,NQ-TABLES,OTT-QA} to more complex, multi-table, and reasoning-intensive tasks~\citep{MMQA,MimoTable}. Nevertheless, practical deployment remains challenging due to the growing size and noise of real-world tables. Traditional retrievers~\citep{bm25,DPR,Colbert} struggle to capture full-table context in token limits, while more recent approaches~\citep{E2E-WTQ,LI-RAGE,TableRAG} primarily rely on table segmentation based on rows, columns, or schema structures to improve matching between query keywords and table fragments. These methods focus on keyword-level matching rather than learning a semantically rich representation of the table that aligns with user questions. Moreover, they are often bound to dataset-specific assumptions or require retriever fine-tuning~\citep{E2E-WTQ,LI-RAGE}.

To tackle this issue, we propose Question Generation from Partial Tables (QGpT), a simple yet effective table retrieval method for long and complex Table-QA tasks. Our approach requires only a small snippet of the table and leverages LLMs to generate simulated questions that are likely to be asked. These questions are jointly embedded with the partial table snippet, enabling a dense representation with minimal token budget while improving retrieval accuracy.

As shown in Figure~\ref{fig:Overview}, QGpT is applied during the offline phase to augment the table corpus. During inference, it can adapt to various retrievers without fine-tuning, and enhances the performance of both dense and late-interaction retrievers~\citep{BGE-m3,Jina-Colbert-v2}. Our QGpT framework offers a lightweight and generalizable solution that improves semantic alignment between questions and tables, reduces context size, and maintains retrieval performance in increasingly large and complex table settings.

\section{Related Work}
\subsection{Table Question Answering (TableQA)}
Early TableQA tasks typically assume the target table is known, focusing on reasoning within a single table. 
Wiki-TableQuestions~\citep{WikiTableQuestions} is a pioneering dataset that provides semi-structured Wikipedia tables with questions involving simple answering. 
WikiSQL~\citep{WikiSQL} reframes the task as a Text-to-SQL problem, enabling structured query-based answering. 
Spider~\citep{spider} introduces higher complexity by requiring reasoning over multiple tables and supporting diverse SQL logic.

Subsequent works began exploring more natural question formulations and open-ended scenarios. 
FeTaQA~\citep{FeTaQA} introduced free-form, multi-hop reasoning questions requiring sentence-level answers across multiple rows. 
OpenWikiTable~\citep{Open-WikiTable} expanded WikiSQL and WikiTableQuestions by incorporating a larger table collection and more natural question expressions.

More recently, MimoTable~\citep{MimoTable} uses real-world spreadsheets of varying size and complexity, containing multi-sheet structures and nested tables. It categorizes difficulty levels based on file count, number of sheets, and header complexity, making it one of the most comprehensive TableQA benchmarks.

Overall, the evolution of TableQA datasets has shifted from “single-table, structured QA pairs” to “multi-table, semantically ambiguous, natural language” tasks. However, most of these datasets do not consider the open-domain setting where tables must be retrieved from a large corpus before answering.

\begin{table*}[t]
  \centering
  \begin{tabular}{llccccc}
    \toprule
    \makecell[c]{\textbf{Retriever}} & \makecell[c]{\textbf{Table size}} & \textbf{R@1} & \textbf{R@3} & \textbf{R@5} & \textbf{R@10} & \textbf{Avg} \\
    \midrule
    \multirow{6}{*}{\centering BGE-m3-dense}
    & Full-Table (8K) & 44.68 {\tiny38.20} & 61.82 {\tiny54.06} & 67.62 {\tiny62.52} & 75.30 {\tiny70.07} & 62.36 {\tiny56.21} \\
    \cmidrule(lr){2-7}
    & 1K tokens & \underline{46.09} {\tiny\underline{40.27}} & \textbf{62.83} {\tiny\textbf{56.22}} & \textbf{68.90} {\tiny\underline{63.47}} & \textbf{77.46} {\tiny{70.06}} & \textbf{63.82} {\tiny{57.51}} \\
    & 2K tokens & 45.15 {\tiny40.15} & \underline{62.75} {\tiny55.62} & \underline{68.16} {\tiny\textbf{63.70}} & \underline{76.70} {\tiny\underline{70.61}} & \underline{63.19} {\tiny\underline{57.52}} \\
    & 5K tokens & 44.53 {\tiny38.86} & 61.82 {\tiny55.26} & 67.69 {\tiny62.69} & 75.53 {\tiny70.32} & 62.39 {\tiny56.78} \\
    \cmidrule(lr){2-7}
    & Top10-rows  & \textbf{46.24} {\tiny\textbf{40.52}} & 62.29 {\tiny\underline{55.99}} & 67.77 {\tiny62.92} & 75.22 {\tiny\textbf{70.98}} & 62.88 {\tiny\textbf{57.60}} \\
    \midrule
    \multirow{6}{*}{\centering Jina-ColBERT-v2}
    & Full-Table (8K) & 42.15 {\tiny26.53} & \underline{57.59} {\tiny37.97} & 61.93 {\tiny43.67} & 69.06 {\tiny54.06} & 57.68 {\tiny40.56} \\
    \cmidrule(lr){2-7}
    & 1K tokens & \underline{48.82} {\tiny\textbf{37.56}} & \underline{64.24} {\tiny\textbf{53.11}} & \textbf{70.40} {\tiny\textbf{58.61}} & \textbf{75.25} {\tiny\textbf{67.09}} & \underline{64.68} {\tiny\textbf{54.09}} \\
    & 2K tokens & 46.67 {\tiny32.09} & 61.10 {\tiny45.69} & 66.22 {\tiny52.44} & 72.10 {\tiny62.24} & 61.52 {\tiny48.12} \\
    & 5K tokens & 42.62 {\tiny26.70} & 56.97 {\tiny39.22} & 62.09 {\tiny44.44} & 68.12 {\tiny54.50} & 57.45 {\tiny41.22} \\
    \cmidrule(lr){2-7}
    & Top10-rows  & \textbf{51.74} {\tiny\underline{36.70}} & \textbf{64.47} {\tiny\underline{51.21}} & \underline{69.23} {\tiny\underline{57.30}} & \underline{74.38} {\tiny\underline{66.11}} & \textbf{64.96} {\tiny\underline{52.83}} \\
    \bottomrule
  \end{tabular}
  \caption{Recall@k performance on the \textbf{MiMoTable-English} dataset (normal font) and \textbf{MiMoTable-Chinese} dataset (shown in \texttt{\tiny next} to each value) across different retrievers and table representation lengths. Note that all table titles in table embeddings are excluded . Best and second-best scores are bolded and underlined respectively per language.}
  \label{tab:Mimotable-table-token-selection}
\end{table*}

\begin{table}[t]
  \centering
  \begin{tabular}{lcc}
    \toprule
    \textbf{Statistics} & \textbf{English} & \textbf{Chinese} \\
    \midrule
    Number of Tables        & 206     & 295     \\
    Number of Sheets        & 295     & 464     \\
    Number of Queries       & 641     & 995     \\
    Max Tokens per Sheet     & 8{,}974 & 1{,}227{,}845 \\
    \midrule
    Tokens / sheet & & \\
    \midrule
    $<$1k             & 35\% & 29\% \\
    1k--2k            & 41\% & 36\% \\
    2k--5k            & 20\% & 22\% \\
    $>$5k             & 4\%   & 13\%  \\
    \bottomrule
  \end{tabular}
  \caption{Comparison of MiMoTable-English and MiMoTable-Chinese statistics.}
  \label{tab:mimotable_stats}
\end{table}

\subsection{Table Retrieval in Open-Domain QA}
A core challenge in open-domain TableQA is efficiently retrieving relevant tables from large corpora. OTT-QA~\citep{OTT-QA} introduces table retrieval from Wikipedia to support open-domain question answering, highlighting the need for integrated retriever-reader systems. NQ-TABLES~\citep{NQ-TABLES} converts questions from Natural Questions~\citep{NQ} into table-based QA pairs and incorporates a retrieval subtask. E2E-WTQ~\citep{E2E-WTQ} extends WikiTableQuestions into a retrieval setting and proposes Cell-Level Table Retrieval (CLTR), focusing on cell-level semantic matching.

LI-RAGE~\citep{LI-RAGE} combines the concept of joinable tables with late-interaction retrievers~\citep{Colbert,Colbertv2}, enabling joint training of retrievers and readers for better performance. 
TableRAG~\citep{TableRAG} decomposes questions and tables into schema-level and cell-level components, enabling compression of million-cell tables into retrievable chunks.

MMQA~\citep{MMQA} introduces Multi-Table Retrieval (MTR), which leverages LLMs like GPT-4~\citep{GPT4} for query decomposition without retriever fine-tuning. It further employs LLM-based retrievers such as TableLlama~\citep{TableLlama} and SGPT~\citep{SGPT}. 

While advances in retriever architectures—from sparse~\citep{bm25} to dense~\citep{DPR,TAPAS} and late-interaction models~\citep{Colbert,Colbertv2,LI-RAGE}—have significantly improved retrieval performance, our proposed framework is agnostic to the underlying retriever. It can flexibly integrate with various retrieval paradigms and consistently enhance performance across different backbone models.

\begin{table*}[t]
  \centering
  \begin{tabular}{llccccc}
    \toprule
    \makecell[c]{\textbf{Retriever}} & \makecell[c]{\textbf{Method}} & \textbf{R@1} & \textbf{R@3} & \textbf{R@5} & \textbf{R@10} & \textbf{Avg} \\
    \midrule
    \multirow{8}{*}{\centering BGE-m3-dense}
    & pT & 44.27 & 61.95 & 67.25 & 75.27 & 62.19 \\
    \cmidrule(lr){2-7}
    & header-only & 33.06 & 49.18 & 57.59 & 65.04 & 51.28  \\
    & desc-only & 36.05 & 51.85 & 61.04 & 71.48 & 55.10 \\
    & QG-only & \underline{48.09} & \underline{64.47} & \textbf{72.39} & \underline{79.28} & \underline{66.06} \\
    \cmidrule(lr){2-7}
    & pT + header & 45.55 & 62.83 & 68.81 & 76.95 & 63.54 \\
    & pT + desc & 45.85 & \underline{64.47} & 71.41 & 78.99 & 65.18\\
    & QGpT & \textbf{50.66} & \textbf{66.42} & \underline{72.35} & \textbf{80.80} & \textbf{67.58} \\
    \midrule
    \multirow{8}{*}{\centering Jina-ColBERT-v2}
    & pT & 54.25 & 69.16 & 75.83 & 81.97 & 70.30 \\
    \cmidrule(lr){2-7}
    & header-only & 51.14 & 67.42 & 74.14 & 80.72 & 68.36 \\
    & desc-only & 47.37 &65.21 &72.04 &81.49 &66.53\\
    & QG-only & 52.76 & 70.10 & 75.17 & 80.45 & 69.62 \\
    \cmidrule(lr){2-7}
    & pT + header & 57.72 & 74.04& \underline{80.08} & \underline{85.34} & \underline{74.29}\\
    & pT + desc & \textbf{58.53} & \underline{74.71} & \textbf{80.40} & \textbf{86.15} & \textbf{74.95} \\
    & QGpT & \underline{57.94} & \textbf{75.21} & 79.11 & 83.26 & 73.88 \\
    \bottomrule
  \end{tabular}
  \caption{
Recall@k results on the \textbf{MiMoTable-English} dataset comparing different table representation strategies and retrievers. The base table input \textit{pT} corresponds to the top-10 rows of each table. Additional representations—headers, descriptions, and questions—are generated using the \textbf{LLaMA3.1-8B-Instruct} model. We evaluate combinations such as \textit{pT} with \textit{header}, \textit{desc} and \textit{QG}. Scores are reported using dense and Multi-vector retriever.}
  \label{tab:Mimotable-table-representation-selection}
\end{table*}

\begin{table}[t]
  \renewcommand{\arraystretch}{1.0}
  \centering
  \small
  \begin{tabular}{lccc}
    \toprule
    \textbf{Dataset} & \textbf{\#Q} & \textbf{\#Tables} & \textbf{Type} \\
    \midrule
    OTT-QA        & 2.2K  & 789         & TQA \\
    FeTaQA        & 2K    & 2K          & TQA \\
    E2E-WTQ       & 241   & 2.1K        & TQA \\
    MiMoTable (en) & 641   & 295 sheets  & Long, TQA \\
    \midrule
    MMQA (2-Tbl)  & 2591  & 2591 / 644  & Long, MTR, \\
    MMQA (3-Tbl)  & 721   & 721 / 391   & TQA, SQL \\
    \bottomrule
  \end{tabular}
  \caption{Dataset statistics. For MMQA, we report both the original and reconstructed table counts (original/ours).}
  \label{tab:data_setting}
\end{table}

\begin{table*}[t]
  \centering
  \begin{tabular}{cccccccc}
    \toprule
    \makecell[c]{\multirow{2}{*}{\textbf{Model}}} & \multicolumn{2}{c}{\multirow{2}{*}{\makecell[c]{\textbf{Method} \\ \textbf{ \& Recall@k}}}} & \multicolumn{4}{c}{\textbf{Dataset}} & \makecell[c]{\multirow{2}{*}{\textbf{Avg}}} \\
    \cmidrule(lr){4-7}
    &  &  & \textbf{MimoTable} & \textbf{OTTQA} & \textbf{FetaQA} & \textbf{E2E-WTQ} & \\
    \midrule
    \multirow{6}{*}{\makecell[c]{BGE-m3\\dense}}
    & pT & \multirow{2}{*}{R@1} & 44.27 & 52.17 & 31.75 & 39.83 & 42.01 \\
    & {\footnotesize \hspace{0.5em}+QGpT} && 50.66 {\tiny↑6.39} & 51.45 {\tiny↓0.72} & 33.95 {\tiny↑2.20} & 41.49 {\tiny↑1.66} & 44.39 {\tiny↑2.38} \\
    \cmidrule(lr){2-8}
    & pT & \multirow{2}{*}{R@5} & 67.25 & 78.27 & 48.08 & 59.75 & 63.34 \\
    & {\footnotesize \hspace{0.5em}+QGpT} && 72.35 {\tiny↑5.10} & 78.14 {\tiny↓0.13} & 50.87 {\tiny↑2.79} & 65.98 {\tiny↑6.23} & 66.84 {\tiny↑3.50} \\
    \cmidrule(lr){2-8}
    & pT & \multirow{2}{*}{R@10} & 75.27 & 86.04 & 55.62 & 70.54 & 71.87 \\
    & {\footnotesize \hspace{0.5em}+QGpT} && 80.80 {\tiny↑5.53} & 86.68 {\tiny↑0.64} & 57.86 {\tiny↑2.24} & 72.61 {\tiny↑2.07} & 74.49 {\tiny↑2.62} \\
    \midrule
    \multirow{6}{*}{\makecell[c]{Jina-\\ColBERT-v2}}
    & pT & \multirow{2}{*}{R@1} & 54.25 & 54.43 & 35.30 & 48.55 & 48.13 \\
    & {\footnotesize \hspace{0.5em}+QGpT} & & 57.94 {\tiny↑3.69} & 55.15 {\tiny↑0.72} & 37.19 {\tiny↑1.89} & 51.45 {\tiny↑2.90} & 50.43 {\tiny↑2.30} \\
    \cmidrule(lr){2-8}
    & pT & \multirow{2}{*}{R@5} & 75.83 & 76.87 & 50.82 & 65.56 & 67.27 \\
    & {\footnotesize \hspace{0.5em}+QGpT} & & 79.11 {\tiny↑3.28} & 78.73 {\tiny↑1.86} & 52.17 {\tiny↑1.35} & 71.37 {\tiny↑5.81} & 70.35 {\tiny↑3.08} \\
    \cmidrule(lr){2-8}
    & pT & \multirow{2}{*}{R@10} & 81.97 & 83.83 & 57.36 & 70.95 & 73.53 \\
    & {\footnotesize \hspace{0.5em}+QGpT} & & 83.26 {\tiny↑1.29} & 86.04 {\tiny↑2.21} & 58.61 {\tiny↑1.25} & 76.76 {\tiny↑5.81} & 76.17 {\tiny↑2.64} \\
    \bottomrule
  \end{tabular}
  \caption{Single-table retrieval performance (Recall@k) across four QA datasets using two retrievers. The base method \textit{pT} uses the top-10 table rows, while \textit{QGpT} denotes the enhancement via question generation. Only embeddings for \textbf{OTTQA} exclude table titles, while all other datasets include them. ↑ indicates the improvement over the corresponding \textit{pT} baseline.}
  \label{tab:Single-table-retrieval}
\end{table*}

\begin{table*}[t]
  \centering
  \begin{tabular}{llccc c}
    \toprule
    \textbf{Retriever} & \textbf{Method} & \textbf{R@2} & \textbf{R@5} & \textbf{R@10} & \textbf{Avg} \\
    \midrule
    \multirow{8}{*}{\centering BGE-m3-dense}
    & pT & 48.65 {\tiny(\underline{39.36})} & 68.54 {\tiny(61.10)} & 80.94 {\tiny(74.69)} & 62.55 \\
    \cmidrule(lr){2-6}
    & MTR (LLaMA3.1-8b) & 44.60 {\tiny(36.50)} & 65.32 {\tiny(60.02)} & 79.07 {\tiny(74.15)} & 59.94 \\
    & {\footnotesize \hspace{1em}+QGpT} & 49.63 {\tiny(38.81)} & 68.33 {\tiny(\underline{62.17})} & 80.13 {\tiny(\textbf{75.86})} & 62.49 \\
    & MTR (GPT4o-mini) & 45.60 {\tiny(36.51)} & 67.32 {\tiny(59.51)} & 79.80 {\tiny(72.61)} & 60.23 \\
    & {\footnotesize \hspace{1em}+QGpT} & {50.19} {\tiny(37.93)} & {71.37} {\tiny(61.24)} & \textbf{82.58} {\tiny(74.86)} & \underline{63.03} \\
    & MTR (GPT4o) & 46.15 {\tiny(36.41)} & 67.79 {\tiny(58.90)} & 81.05 {\tiny(73.17)} & 60.58 \\
    & {\footnotesize \hspace{1em}+QGpT} & \underline{51.10} {\tiny(38.09)} & \textbf{71.89} {\tiny(60.40)} & {82.13} {\tiny(74.45)} & 63.01 \\
    \cmidrule(lr){2-6}
    & QGpT & \textbf{52.24} {\tiny(\textbf{40.05})} & \underline{71.47} {\tiny(\textbf{63.36})} & \underline{82.33} {\tiny(\underline{75.36})} & \textbf{64.14} \\
    \midrule
    \multirow{8}{*}{\centering Jina-ColBERT-v2}
    & pT & \underline{58.18} {\tiny(\underline{45.79})} & \underline{77.87} {\tiny(\underline{70.16})} & \underline{87.09} {\tiny(\underline{81.63})} & \underline{70.12} \\
    \cmidrule(lr){2-6}
    & MTR (LLaMA3.1-8b) & 54.34 {\tiny(42.65)} & 73.90 {\tiny(66.84)} & 83.93 {\tiny(78.30)} & 66.66 \\
    & {\footnotesize \hspace{1em}+QGpT} & 54.28 {\tiny(43.52)} & 74.98 {\tiny(67.67)} & 84.22 {\tiny(79.71)} & 67.40 \\
    & MTR (GPT4o-mini) & 56.40 {\tiny(41.24)} & 75.27 {\tiny(65.18)} & 86.36 {\tiny(75.96)} & 66.74 \\
    & {\footnotesize \hspace{1em}+QGpT} & 56.33 {\tiny(42.15)} & 76.18 {\tiny(66.81)} & 86.17 {\tiny(78.11)} & 67.63 \\
    & MTR (GPT4o) & 56.76 {\tiny(42.15)} & 75.82 {\tiny(65.72)} & 86.80 {\tiny(76.53)} & 67.30 \\
    & {\footnotesize \hspace{1em}+QGpT} & 56.99 {\tiny(43.23)} & 76.91 {\tiny(67.82)} & 86.91 {\tiny(77.97)} & 68.31 \\
    \cmidrule(lr){2-6}
    & QGpT & \textbf{59.49} {\tiny(\textbf{46.75})} & \textbf{78.41} {\tiny(\textbf{71.43})} & \textbf{87.25} {\tiny(\textbf{83.14})} & \textbf{71.08} \\
    \bottomrule
  \end{tabular}
  \caption{Recall@k performance on the \textbf{MMQA} dataset with different retrievers and query methods. Metrics reflect performance on the 2-table setting; scores in parentheses show corresponding results under the 3-table setting (in {\tiny small font}). All methods are built upon the \textit{pT} baseline (top-10 table rows). \textit{MTR} uses sub-query generation via \textbf{LLaMA3.1-8B-Instruct}, \textbf{GPT-4o}, or \textbf{GPT-4o-mini}. \textit{QGpT} questions are generated using \textbf{LLaMA3.1-8B-Instruct}.}
  \label{tab:Multi-Table-Retrieval}
\end{table*}

\begin{table*}[t]
  \centering
  \begin{tabular}{lccccc}
    \toprule
    \makecell[c]{\textbf{Model \& Method}} & \textbf{R@1} & \textbf{R@3} & \textbf{R@5} & \textbf{R@10} & \textbf{Avg} \\
    \midrule
    BGE-m3-dense {\footnotesize {(pT)}} & 52.17 & 70.73 & 78.27 & 86.04 & 71.80 \\
    {\footnotesize \hspace{1em} +QGpT w/o title} 
    & 51.45 {\tiny↓0.72} 
    & 70.64 {\tiny↓0.09} 
    & 78.14 {\tiny↓0.13} 
    & 86.68 {\tiny↑0.64} 
    & 71.23 {\tiny↓0.57} \\
    {\footnotesize \hspace{1em} +QGpT w/ title} 
    & 60.79 {\tiny↑8.62} 
    & 78.14 {\tiny↑7.41} 
    & 84.37 {\tiny↑6.10} 
    & 91.46 {\tiny↑5.42} 
    & 78.69 {\tiny↑6.89} \\
    \cmidrule(lr){1-6}
    Jina-ColBERT-v2 {\footnotesize {(pT)}} & 54.43 & 70.01 & 76.87 & 83.83 & 71.29 \\
    {\footnotesize \hspace{1em} +QGpT w/o title} 
    & 55.15 {\tiny↑0.72} 
    & 71.27 {\tiny↑1.26} 
    & 78.73 {\tiny↑1.86} 
    & 86.04 {\tiny↑2.21} 
    & 72.80 {\tiny↑1.51} \\
    {\footnotesize \hspace{1em} +QGpT w/ title} 
    & 60.07 {\tiny↑5.64} 
    & 75.34 {\tiny↑5.33} 
    & 80.80 {\tiny↑3.93} 
    & 87.71 {\tiny↑3.88} 
    & 75.98 {\tiny↑4.70} \\
    \bottomrule
  \end{tabular}
  \caption{Recall@k results on the \textbf{OTT-QA} dataset with different retrievers and \textit{QGpT} enhancements. All embedding representations exclude table titles, while \textit{QGpT} question generation is conducted \textbf{with or without referencing table titles}.}
  \label{tab:OTTQA-title-ablation}
\end{table*}
\section{Methodology}
\subsection{Partial Table Selection}
To support complex TableQA tasks, it is essential to reduce large tables by removing irrelevant cells. A common approach is to retain only the header / schema or to constrain the input by a fixed number of tokens or rows. However, the former is heavily dependent on the nature of the dataset: SQL-based datasets typically rely heavily on headers, while in datasets like MiMoTable~\citep{MimoTable}, which aim to increase difficulty, headers are not necessarily positioned in the first row and may even be multi-level.

To inform our strategy for partial table selection, we first analyze the length distribution of tables in the MiMoTable dataset (see Table~\ref{tab:mimotable_stats}). Given the wide variance in table sizes, we compare two representative approaches for truncation: limiting by token length and selecting the top-10 rows. We construct English and Chinese table corpora accordingly and evaluate retrieval performance using BGE-m3~\citep{BGE-m3} and Jina-ColBERT-v2~\citep{Jina-Colbert-v2}—both supporting up to 8K tokens—as our retrievers. As shown in Table~\ref{tab:Mimotable-table-token-selection}, the top-10 rows selection achieves comparable performance to 1K tokens truncation in complex table QA tasks. Considering the diversity of table lengths across datasets and the simplicity of implementation, we adopt the top-10 rows as our default strategy for partial table input.

To further improve semantic alignment between user questions and compressed table inputs, we explore several strategies for enriching table representations. As illustrated in Figure~\ref{fig:Table-representation}, these include selecting table headers, generating natural language descriptions, and producing synthetic questions by LLMs. To realize these strategies, we leverage LLaMA-3.1-8B-Instruct~\citep{LLAMA8b} to implement these strategies. Table~\ref{tab:Mimotable-table-representation-selection} presents the retrieval results using each method. Interestingly, even without table content, simulated questions alone can achieve comparable or better performance than partial tables. Combining simulated questions with partial tables further improves retrieval performance consistently. Motivated by these findings, we propose a unified table representation method—\textbf{QGpT} (Question Generation from Partial Tables)—which augments the top-10 table rows with generated questions to construct an enriched table corpus.

\subsection{Question Generation from Partial Tables}

QGpT is highly extensible and model-agnostic. The simulated questions are generated during an offline preprocessing stage, enabling integration with various retrieval paradigms (e.g., query decomposition or different retrievers) during online inference.

\paragraph{Offline Stage}  
Given a table corpus $\mathcal{C}_T$, we convert each table into a markdown format and extract its name and the top-10 rows to construct a new partial table corpus $\mathcal{C}_t$. For each partial table $t_i \in \mathcal{C}_t$, we use a language model $M$ to generate a set of questions $\{q_{t_i,j}\}$ such that the number of generated questions $j$ satisfies:
\[
j \geq \left\lceil \frac{|\mathcal{H}_{t_i}|}{2} \right\rceil
\]
where $|\mathcal{H}_{t_i}|$ is the number of headers in Table $t_i$. The resulting \textit{augmented} partial table $t'_i = (t_i, q_{t_i,1}, q_{t_i,2}, \ldots, q_{t_i,j})$ is then embedded using an embedding model $E$, producing a set of vectors for the table corpus:
\[
E(\mathcal{C}_{t'}) = \{ E(t'_1), E(t'_2), \ldots, E(t'_n) \}
\]

\paragraph{Online Stage}  
Given a user query $q_i \in \mathcal{Q}$, we compute its embedding $E(q_i)$ and perform cosine similarity with all table representations:
\[
\text{sim}(E(q_i), E(t'_j)) \quad \forall j \in [1, n]
\]
The top-$k$ most similar entries are then retrieved.

\section{Experiments}
\subsection{Experimental Settings}



\paragraph{Datasets} For the Single Table Retrieval task, we conduct experiments on OTT-QA~\citep{OTT-QA} and FeTaQA~\citep{FeTaQA} following the evaluation settings proposed by TARGET~\citep{Target}, as well as on E2E-WTQ~\citep{E2E-WTQ} and MiMoTable-English~\citep{MimoTable}.

For the Multi-Table Retrieval task, we evaluate on the MMQA dataset~\citep{MMQA}. However, the original MMQA dataset only provides question-table-answer triples without releasing a unified table corpus. Moreover, tables with the same \texttt{table\_name} may differ in structure across different examples—some pointing to semantically distinct tables despite identical names. To ensure consistency and reproducibility, we reconstruct the MMQA table corpus by performing a schema-based deduplication. Specifically, we (1) group tables by \texttt{table\_name} and enumerate distinct schema variants, (2) assign unique identifiers (e.g., \texttt{department\_\_1}, \texttt{department\_\_2}) for structurally distinct tables, and (3) update all question-table mappings accordingly. This process results in a flattened and de-duplicated table corpus that supports robust retrieval evaluation. A summary of all datasets used can be found in Table~\ref{tab:data_setting}.

\paragraph{Models} Throughout all experiments, we adopt BGE-m3~\citep{BGE-m3} and Jina-ColBERT-v2~\citep{Jina-Colbert-v2} as base retrievers, which support up to 8K tokens and represent dense and late-interaction paradigms, respectively. All question generation is performed using LLaMA-3.1-8B-Instruct~\citep{LLAMA8b}, with prompt details provided in Appendix~\ref{sec:appendix_prompt}. 

To analyze the impact of model capacity on query decomposition in Multi-Table Retrieval (MTR), we experiment with multiple large language models, including LLaMA-3.1-8B-Instruct, GPT-4o-mini, and GPT-4o~\citep{GPT4o}, within the MTR framework.

\paragraph{Evaluation Metrics} We report performance using the \textbf{Recall@\(\mathbf{K}\)}
 metric across all experiments.
For Multi-Table Retrieval, we follow the MMQA evaluation settings with $K = \{2, 5, 10\}$.

\subsection{Implementation Details}
All experiments are conducted using two NVIDIA RTX A6000 GPUs. We use RAGatouille\footnote{\url{https://github.com/AnswerDotAI/RAGatouille}} and Milvus\footnote{\url{https://milvus.io/}}~\citep{Milvus} as the retrieval infrastructure for vector indexing and search for Jina-ColBERT-v2 and BGE-m3, respectively. For Milvus, we adopt \texttt{index\_type=IVF\_FLAT}, \texttt{metric\_type=IP}, and \texttt{nlist=256} to balance retrieval speed and accuracy. For RAGatouille, we leverage PLAID~\citep{PLAID}, a high-performance indexing engine specifically designed for late interaction retrievers, enabling efficient token-level matching at scale.

\section{Experimental Results}
\subsection{Main Results}

\paragraph{Single Table Retrieval} We evaluate retrieval performance across datasets ranging from short and simple to long and complex tables. As shown in Table~\ref{tab:Single-table-retrieval}, QGpT consistently improves retrieval performance across all Single Table QA datasets compared to using partial tables alone. Notably, on the MiMoTable dataset—which features longer and more complex tables—both the dense and late-interaction retrievers benefit significantly from QGpT. This demonstrates that QGpT is particularly effective in scenarios where aggressive table compression is required, offering robust performance regardless of table complexity.

\paragraph{Multi Table Retrieval} Building on the same partial table setup, we compare Multi-Table Retrieval (MTR), QGpT, and their combination on the MMQA dataset. Note that our evaluation is based on a reconstructed table corpus. Therefore, our results are not directly comparable to those reported in the original paper.

Table~\ref{tab:Multi-Table-Retrieval} shows that while MTR performance slightly improves with larger query decomposition models (e.g., GPT-4o), it still underperforms compared to directly using partial tables for retrieval. This gap may stem from differences in implementation, such as our exclusion of the original paper's hand-crafted one-shot examples or the use of full tables during retrieval. However, integrating QGpT into MTR substantially closes the performance gap, and QGpT alone consistently outperforms the baseline across settings. These results highlight QGpT’s effectiveness in both single- and multi-table retrieval, providing strong gains even when only limited table information is available.

\subsection{Ablation studies}
\paragraph{Can Simulated Questions Bridge the Semantic Gap?} To understand whether simulated questions truly enhance the semantic alignment between partial tables and questions, we conduct an ablation study on the OTT-QA dataset. As noted by TARGET~\citep{Target}, OTT-QA questions often align closely with table titles, and excluding titles from the corpus can dramatically reduce BM25 Recall@$10$ from $95\%$ to $44\%$.

In our study, we generate two versions of simulated questions using QG models: one with access to table titles, and another without. We then embed both variants alongside partial tables that exclude the titles. As shown in Table~\ref{tab:OTTQA-title-ablation}, incorporating simulated questions generated with access to titles leads to significant performance improvements, outperforming both the baseline and the QGpT variant that does not use titles. These results validate that simulated questions can effectively bridge the semantic gap, helping partial tables better align with user queries during retrieval.

\section{Conclusion}
In this work, we propose QGpT (Question Generation from Partial Tables), a simple yet effective framework to enhance table retrieval by bridging the semantic gap between compressed table inputs and user queries. By leveraging LLMs to generate simulated questions from partial tables, QGpT provides a semantically enriched representation that improves retrieval performance across both single- and multi-table QA benchmarks.

Our extensive experiments demonstrate that QGpT consistently outperforms traditional partial table baselines on diverse datasets, including long, noisy, and multi-table settings. Notably, the framework is model-agnostic and can be flexibly integrated with different retrievers without requiring fine-tuning.

\section*{Limitations}
While QGpT offers a generalizable solution for enhancing table retrieval, several limitations remain:

\textbf{LLM dependency:} The quality of simulated questions relies heavily on the capabilities of the underlying LLM. Lower-quality LLMs may generate irrelevant or redundant questions, limiting retrieval gains.

\textbf{Generation latency:} Although question generation occurs offline, large-scale preprocessing for hundreds of thousands of tables may introduce overhead in real-world deployments.

\section*{Acknowledgement}
This work is supported by NSTC 112-2634-F-005-002-project Smart Sustainable New Agriculture Research Center (SMARTer), NSTC Taiwan Project under grant 112-2221-E-005-075-MY3, and Delta Research Center.

\bibliography{custom}
\clearpage
\appendix
\section{Prompt Details for Question Generation} \label{sec:appendix_prompt}

We provide the exact prompts used in our LLM-based pipeline for extracting headers and generating simulated questions from partial tables. Prompts were designed to be highly structured and instructive, guiding the LLM (LLaMA-3.1-8B-Instruct) to handle inputs.
All prompts return a strict \texttt{JSON} format, suitable for programmatic post-processing.

\subsection{Header Extraction + Question Generation (Full Pipeline)} If header extraction is desired (used in most QGpT scenarios), we use the following prompt format:

\begin{quote} \small \ttfamily You are an expert in table data analysis. Given a table with its file name, sheet name, and a portion of its content (first ten rows), your task is to **extract key headers and generate questions** based on the table \& headers.

\textbf{Important Considerations:} \begin{itemize} \item The table may contain nan or Unnamed: values, which represent empty merged cells in the original table. These **should not** be considered as meaningful data points or headers. \item The **true column headers may not always be in the first row or first column**. Carefully analyze the table to identify the correct headers. \item If the table has **multi-level headers**, preserve the hierarchical structure without merging or altering the text. \item If the table has an **irregular header structure** (such as key-value formatted headers where column names are listed separately), extract the correct header names accordingly. \item **Ignore rows that contain mostly empty values (nan, Unnamed:) or placeholders without meaningful data.** \item **Do not generate python code, extract headers and questions on your own.** \item The type of Questions could be one of (lookup, calculate, visualize, reasoning).  \item **Generate question using the language of the table.** \end{itemize}

\textbf{Tasks:} \begin{enumerate} \item \textbf{Extract Header Names:} \begin{itemize} \item Identify the **true headers** by analyzing the structure of the table. \item **Exclude** placeholder values like "nan" and "Unnamed:". \item If the table contains **multi-level headers**, keep them as separate levels without merging. \item If the table has **key-value headers**, extract the correct column names. \end{itemize} \item \textbf{Generate Questions (Context-Specific to the Table):} \begin{itemize} \item Formulate **questions that can only be answered using this specific table**. \item  Ensure **each question involves 1 to 3 different headers** to capture interactions between data \& columns. \item Ensure the header diversity in all the questions. \item Use '' to mark the headers in the question. \item **Total number of questions should larger than the half number of extracted headers** \end{itemize} \end{enumerate}

\textbf{**Output Format (Strictly JSON format)**} Only return a JSON dictionary object with the extracted headers and questions, without any additional explanations or formatting. \begin{verbatim} {{ "headers": ["header1", "header2", \end{verbatim}
\begin{verbatim} "..."], "questions": ["question1", \end{verbatim}
\begin{verbatim} "question2", "..."] }} \end{verbatim}

\textbf{Input Table:}
\texttt{<{table}>} \end{quote}

\subsection{Question Generation Only (Without Header Extraction)} If header extraction is skipped (e.g., MMQA), we apply a simplified prompt:

\begin{quote} \small \ttfamily You are an expert in table data analysis. Given a table with its file name and a portion of its content (first ten rows), your task is to **generate questions** based on the table \& headers.

\textbf{Important Considerations:} \begin{itemize} \item **Do not generate python code, generate questions on your own.** \item The type of Questions could be one of (Numerical, List, Count, Select). \item **Generate question using the language of the table.** \end{itemize}

\textbf{**Tasks:**} \begin{itemize} \item **1. Generate Questions (Context-Specific to the Table):** \item Formulate **questions that can only be answered using this specific table**. \item Ensure **each question involves 1 to 3 different headers** to capture interactions between data \& columns. \item Ensure the header diversity in all the questions. \item Use '' to mark the headers in the question. \item **Total number of questions should larger than the half number of extracted headers** \end{itemize}

\textbf{**Output Format (Strictly JSON format)**} Only return a JSON dictionary object with the extracted headers and questions, without any additional explanations or formatting.\begin{verbatim} { "questions": ["question1",\end{verbatim}
\begin{verbatim} "question2","..."]\end{verbatim}

\textbf{Input Table:}
\texttt{<{table}>} \end{quote}

\nocite{}
\end{document}